\documentclass[twocolumn]{article}
\usepackage[utf8]{inputenc}
\usepackage[margin=0.5in]{geometry}
\usepackage{graphicx}
\usepackage[none]{hyphenat}
\usepackage{amssymb}
\usepackage{amsmath}
\usepackage{amsthm}
\usepackage{physics}
\usepackage{mathrsfs}
\usepackage{hyperref}
\usepackage{bold-extra}
\usepackage{xcolor}
\usepackage{pdfrender,xcolor}

\newcommand{\ts}{\textsuperscript}
\newlength\myindent
\begin{document}
\pdfrender{StrokeColor=black,TextRenderingMode=2,LineWidth=0.5pt}
\emergencystretch 1em
\begin{center}
\large{\textbf{MATHEMATICAL MODELLING AND A NUMERICAL SOLUTION FOR HIGH PRECISION SATELLITE EPHEMERIS DETERMINATION}}\\$\,$\\
\small Aravind Gundakaram$^{\dagger}$, Abhirath Sangala$^{\dagger}$, Aditya Sai Ellendula$^{*}$, Prachi Kansal$^{*}$, Lanii Lakshitaa$^{*}$, Suchir Reddy Punuru$^{*}$, Nethra Naveen$^{\dagger}$, Sanjitha Jaggumantri$^{*}$
\rule{\columnwidth}{0.1pt}
\footnotesize\par $\dagger$ \emph{The authors are students of the Department of Mathematics, Mahindra University, Hyderabad-$500043$, Telangana, India.}\par $*$ \emph{The authors are students of the Department of Computer Science, Mahindra University, Hyderabad-$500043$, Telangana, India.}\par\emph{The authors have received the generous help of the faculty of Mahindra University in completing the work and finalising the manuscript. The authors are also indebted to ISRO (Bhuvan) and CelesTrak for providing open source data sets to validate the work.}\par\emph{The authors express their gratitude to Mahindra University for financial assistance in attending the $9^{th}$ International Conference and Exhibition on Satellite and Space Missions on September $11-12$, $2023$ in Toronto, Canada.}\\
\rule{\columnwidth}{0.1pt}
\end{center}
\normalsize\centerline{\textbf{ABSTRACT}}
\small\par In this paper, we develop a high-precision satellite orbit determination model for satellites orbiting the Earth. Solving this model entails numerically integrating the differential equation of motion governing a two-body system, employing Fehlberg's formulation and the Runge-Kutta class of embedded integrators with adaptive stepsize control. Relevant primary perturbing forces included in this mathematical model are the full force gravitational field model, Earth's atmospheric drag, third body gravitational effects and solar radiation pressure. Development of the high-precision model required accounting for the perturbing influences of Earth radiation pressure, Earth tides and relativistic effects.\par The model is then implemented to obtain a high-fidelity Earth orbiting satellite propagator, namely the Satellite Ephemeris Determiner (SED), which is comparable to the popular High Precision Orbit Propagator (HPOP). The architecture of SED, the methodology employed, and the numerical results obtained are presented.\\\\
\normalsize\centerline{\textbf{\S1. INTRODUCTION}}
\small\par The interest in developing satellite orbit propagators has been driven by the primary requirement of predicting the future locations of the several satellites orbiting the Earth. Analytical, semi-analytical, and numerical propagators have been developed to approach this problem. The analytical propagators, while computationally efficient, have the disadvantage of being limited in their accuracy due to local truncation errors. The special perturbation methods overcome this limitation, and are accurate for short term propagation. However, during long term propagation, these methods have the drawback of accumulating high round-off errors. They are also computationally demanding. Semi-analytical propagators were developed to achieve a reasonable trade-off between precision and computational effort.$^{[1]}$\par Numerical orbit propagators consist of the differential equation governing motion in orbit, the mathematical model that describes the environment, and the numerical scheme that acts as the integrator. In this paper, we consider the two-body problem governed by Kepler's differential equation of motion. This unperturbed model is solved using the Runge-Kutta-Fehlberg class of adaptive step controlled numerical integrators, the embedded RKF-$4(5)$, RKF-$7(8)$, and RKF-$8(9)$ methods. They have the advantage of being explicit, stable, and easy to implement.\par Perturbations acting on the satellite may be categorised into conservative and non conservative forces. The force model considered in this paper includes the perturbative effects from the gravity potential field, atmospheric drag, solar and lunar gravitational accelerations, solar radiation pressure, Earth radiation pressure, Earth tides, and relativistic effects. The influence of these forces varies depending on the altitude and the physical dimensions of the satellite.$^{[2,3]}$\par The current work presents explicit expressions to calculate the force exerted by each perturbation contributing to the total acceleration of the satellite. Deducing these expressions sometimes involves introducing appropriate approximations. In situations where calculating certain terms within these expressions poses practical challenges, we have provided recurrence relations to assist with implementation. The resultant model and its implementation yield a high-fidelity satellite orbit propagator, which we call the Satellite Ephemeris Determiner (SED). The results obtained from SED are compared with the widely used High Precision Orbit Propagator (HPOP). Currently, SED is implemented for the two-body problem. Further work will also enable it to handle the three-body problem.\par The selection of the frame of reference is central to high-precision models. We work in the Earth-Centered Inertial (ECI) J$2000$ frame of reference. The rest of the paper is divided into sections $2$ through $7$. Section $2$ describes the basic mathematical model involving only dominant perturbing forces, while section $3$ introduces some additional forces only necessary for high-precision modelling. Section $4$ details the numerical scheme to solve the high-precision perturbed model. Section $5$ describes the computational methodology used in the implementation of the model and section $6$ discusses results obtained from the implementation. Finally, section $7$ summarises our conclusions.\\\\
\normalsize\centerline{\textbf{\S2. THE MATHEMATICAL MODEL}}
\small\par The unperturbed Newtonian two-body differential equation of motion in an inertial reference frame is expressed by \[\frac{d^2\vec r}{dt^2}=\frac{-\mu}{r^3}\vec r,\tag{1}\] where $\vec r$ is the satellite's geocentric position vector at time $t$, with magnitude $r$, and $\mu$ is Earth's gravitational constant. This equation assumes that all of the Earth's mass is concentrated at the center of the coordinate system, and the law of gravitation $\ddot{\vec r}=(-GM_\oplus/r^3)\vec r$ can be used to compute the rate of acceleration experienced by the satellite at distance vector $\vec r$. In the subsequent discussion concerning a more practical representation, it is advantageous to utilize the gradient of the corresponding gravitational potential $U$, such that $\ddot{\vec r}=\nabla U$, where $U=GM_\oplus/\vec r.$ This formulation for the potential can be readily extended to any arbitrary mass distribution by aggregating the effects generated by each individual mass component $dm=\rho(s)\,d^3s$, in accordance with$^{[4]}$ \[U=G\int\frac{\rho(s)\,d^3s}{|\vec r-\vec s|},\tag{2}\] where, $\rho(s)$ represents the density at a certain point $\vec s$ within the Earth, and $|\vec r-\vec s|$ is the distance of the satellite from $\vec s$. Evaluating the integral in $(2)$ involves expanding the reciprocal of the distance $|\vec r-\vec s|$ as a series of Legendre polynomials. For $r>s$, we have \[\frac{1}{|\vec r-\vec s|}=\frac{1}{r}\sum\limits_{n=0}^\infty\left(\frac{s}{r}\right)^nP_n(\cos\gamma),\tag{3}\]\[\text{where}\quad P_n(x)=\frac{1}{2^nn!}\frac{d^n}{dx^n}(x^2-1)^n\tag{4}\] is the Legendre polynomial of degree $n$ and $\gamma$ represents the angle between $\vec r$ and $\vec s$. We calculate the latitude $\phi$ and longitude $\lambda$ measured with respect to the Earth's center for the point $\vec r$ in accordance with $x=r\cos\phi\cos\lambda$, $y=r\cos\phi\sin\lambda$, and $z=r\sin\phi$. Together with the corresponding values $\phi'$ and $\lambda'$ for $\vec s$, we can utilize the addition theorem for Legendre polynomials, which states that$^{[5]}$ \[P_n(\cos\gamma)=\left(\sum\limits_{m=0}^n(2-\delta_{0m})\frac{(n-m)!}{(n+m)!}P_{nm}(\sin\phi)\right)\zeta\,,\tag{5}\]\[\text{where\quad}\zeta=P_{nm}(sin\phi')\cos(m(\lambda-\lambda')).\] $P_{nm}$ here is the associated Legendre polynomial of degree $n$ and order $m$, defined by \[P_{nm}(x)=(1-x^2)^{m/2}\frac{d^m}{dx^m}P_n(x).\tag{6}\] At this stage, we can express the gravitational potential of the Earth in the form$^{[1,3]}$ \[U=\frac{GM_\oplus}{r}\sum\limits_{n=0}^\infty\sum\limits_{m=0}^n\frac{R_\oplus^n}{r^n}P_{nm}(\sin\phi)(C_{nm}\cos(m\lambda)+S_{nm}\sin(m\lambda)),\tag{7}\]with the coefficients\[\quad C_{nm}=\frac{2-\delta_{0m}}{M_\oplus}\frac{(n-m)!}{(n+m)!}\int\frac{s^n}{R^n_\oplus}P_{nm}(\sin\phi')\cos(m\lambda')\rho(s)d^3s,\tag{8}\]\[S_{nm}=\frac{2-\delta_{0m}}{M_\oplus}\frac{(n-m)!}{(n+m)!}\int\frac{s^n}{R^n_\oplus}P_{nm}(\sin\phi')\sin(m\lambda')\rho(s)d^3s,\tag{9}\] that describe the dependence on Earth's internal mass distribution. To account for the wide range of magnitudes exhibited by the geopotential coefficients $C_{nm}$ and $S_{nm}$, the normalised coefficients $\bar{C}_{nm}$ and $\bar{S}_{nm}$ are used. These are defined by the equation \[\begin{bmatrix}\bar C_{nm}\\\bar S_{nm}\end{bmatrix}=\sqrt{\frac{(n+m)!}{(2-\delta_{0m})(2n+1)(n-m)!}}\begin{bmatrix}C_{nm}\\S_{nm}\end{bmatrix}.\tag{10}\] The normalized coefficients demonstrate a greater consistency in their magnitudes, and their approximate sizes are determined by the empirical rule $\bar C_{nm},\bar S_{nm}\approx10^{-5}/n^2$. Thus, the gravitational acceleration resulting from Earth's gravity potential may be written as \[\frac{d^2\vec r}{dt^2}=\nabla\frac{GM_\oplus}{r}\sum\limits_{n=0}^\infty\sum\limits_{m=0}^n\frac{R_\oplus^n}{r^n}\bar P_{nm}(\sin\phi)\qquad\qquad\qquad\]\[\cdot(\bar C_{nm}\cos(m\lambda)+\bar S_{nm}\sin(m\lambda)),\tag{11}\] where the normalised associated Legendre polynomials are given by$^{[5]}$ \[\bar P_{nm}=\sqrt{\frac{(2-\delta_{0m})(2n+1)(n-m)!}{(n+m)!}}P_{nm}.\tag{12}\] When calculating the gravitational potential of the Earth at a specific point, multiple recurrence relations can be employed to evaluate the Legendre polynomials. Starting with $P_{00}=1$, all polynomials $P_{mm}$ up to the intended degree and order can be computed using $P_{mm}(x)=(2m-1)(1-x^2)^{1/2}P_{m-1,m-1}$. Using these results, the remaining polynomials can be evaluated using $P_{m+1,m}(x)=(2m+1)xP_{mm}(x)$ and \[P_{nm}(x)=\frac{1}{n-m}((2n-1)xP_{n-1,m}(x)\qquad\qquad\qquad\qquad\quad\]\[-(n+m-1)P_{n-2,m}(x))\,\,\forall\,n>m+1.\tag{13}\] 
This enables us to compute the gravitational potential $(7)$ and the resulting acceleration $(11)$ as a function of the Cartesian coordinates $(x, y, z)$ of the satellite. Defining \[V_{nm}=\left(\frac{R_\oplus}{r}\right)^{n+1}P_{nm}(\sin\phi)\cos(m\lambda),\]\[ W_{nm}=\left(\frac{R_\oplus}{r}\right)^{n+1}P_{nm}(\sin\phi)\sin(m\lambda),\tag{14}\] the gravitational potential can be expressed as$^{[1,3]}$ \[U=\frac{GM_\oplus}{R_\oplus}\sum\limits_{n=0}^\infty\sum\limits_{m=0}^n(C_{nm}V_{nm}+S_{nm}W_{nm}).\tag{15}\] The $V_{nm}$ and $W_{nm}$ additionally fulfil the subsequent recurrence relations for all values $n\geq m+1$. \[V_{mm}=(2m-1)\left(\frac{xR_\oplus}{r^2}V_{m-1,m-1}-\frac{yR_\oplus}{r^2}W_{m-1,m-1}\right),\]\[W_{mm}=(2m-1)\left(\frac{xR_\oplus}{r^2}W_{m-1,m-1}+\frac{yR_\oplus}{r^2}V_{m-1,m-1}\right),\tag{16}\]\[V_{nm}=\left(\frac{2n-1}{n+m}\right)\frac{zR_\oplus}{r^2}V_{n-1,m}+\left(\frac{n+m-1}{n-m}\right)\frac{R_\oplus^2}{r^2}V_{n-2,m},\]\[W_{nm}=\left(\frac{2n-1}{n+m}\right)\frac{zR_\oplus}{r^2}W_{n-1,m}+\left(\frac{n+m-1}{n-m}\right)\frac{R_\oplus^2}{r^2}W_{n-2,m}.\tag{17}\] Furthermore, $V_{00}=R_\oplus/r$ and $W_{00}=0$ are known. To calculate all the $V_{nm}$ and the $W_{nm}$, we first obtain all the terms $V_{n0}$ by using $(17)$. The corresponding $W_{n0}$ are all identically equal to zero. $(16)$ then yields the terms $V_{11}$ and $W_{11}$ from $V_{00}$, and the determination of all the $V_{n1}$ follows. The $V_{nm}$ and the $W_{nm}$ may now be recursively calculated. The acceleration $\ddot{\vec r}$ can now be computed from the $V_{nm}$ and the $W_{nm}$ as \[\frac{d^2x}{dt^2}=\sum\limits_{n,m}\frac{d^2x_{nm}}{dt^2},\quad\frac{d^2y}{dt^2}=\sum\limits_{n,m}\frac{d^2y_{nm}}{dt^2},\quad\frac{d^2z}{dt^2}=\sum\limits_{n,m}\frac{d^2z_{nm}}{dt^2}.\tag{18}\] The component accelerations $\ddot x_{nm}$, $\ddot y_{nm}$, and $\ddot z_{nm}$ are given by the following equalities.$^{[1]}$ \[\frac{d^2x_{nm}}{dt^2}\overset{(m=0)}{=}\frac{GM}{R_\oplus^2}(-C_{n0}V_{n-1,1}),\]\[\frac{d^2x_{nm}}{dt^2}\overset{(m>0)}{=}\frac{GM}{R_\oplus^2}\frac{1}{2}\Bigl((-C_{nm}V_{n+1,m+1}-S_{nm}W_{n+1,m+1})\qquad\qquad\qquad\qquad\]\[+\frac{(n-m+2)!}{(n-m)!}(C_{nm}V_{n+1,m-1}+S_{nm}W_{n+1,m-1})\Bigr),\tag{19}\]\[\frac{d^2y_{nm}}{dt^2}\overset{(m=0)}{=}\frac{GM}{R_\oplus^2}(-C_{n0}W_{n+1,1}),\]\[\frac{d^2y_{nm}}{dt^2}\overset{(m>0)}{=}\frac{GM}{R_\oplus^2}\frac{1}{2}\Bigl((-C_{nm}W_{n+1,m+1}+S_{nm}V_{n+1,m+1})\qquad\qquad\qquad\qquad\]\[+\frac{(n-m+2)!}{(n-m)!}(-C_{nm}W_{n+1,m-1}+S_{nm}V_{n+1,m-1})\Bigr),\tag{20}\]\[\frac{d^2z_{nm}}{dt^2}=\frac{GM}{R_\oplus^2}(n-m+1)(-C_{nm}V_{n+1,m}-S_{nm}W_{n+1,m}).\tag{21}\] Gravitational coefficients are determined through experimental means, involving the data analysis of artificial satellites, and through gravimetric methods. In this paper, we choose the values of the normalised coefficients according to the Joint Gravity Model (JGM-$3$).$^{[6,7]}$\par We now consider the perturbing acceleration due to solar and lunar gravitational influences. In accordance with Newton's law of gravitation, the acceleration of a satellite caused by a point mass $M$ is determined by \[\frac{d^2\vec r}{dt^2}=GM\frac{\vec s-\vec r}{|\vec s-\vec r|^3},\tag{22}\] where $\vec s$ and $\vec r$ are the geocentric position vectors of the mass $M$ and the satellite respectively. But this is with respect to an inertial frame of reference, whereas the Earth itself is subject to the acceleration \[\frac{d^2\vec r}{dt^2}=GM\frac{\vec s}{|\vec s|^3}\tag{23}\] due to the mass $M$. $(23)$ must be subtracted from $(22)$ to obtain the acceleration of the satellite in the Earth centered frame of reference. For Earth orbiting satellites, the dominating masses that act as the mass $M$ are either the Sun or the Moon. \[\frac{d^2\vec r}{dt^2}=GM\left(\frac{\vec s-\vec r}{|\vec s-\vec r|^3}-\frac{\vec s}{|\vec s|^3}\right).\tag{24}\] It is not necessary to know the coordinates of the Sun and the Moon to the highest precision while calculating their gravitational influence on an Earth orbiting satellite, because the said influence is much smaller as compared to the central attraction of the Earth. However, if there is a frequent need for precise numerical coordinates of the Sun and the Moon, approximations using \v{C}eby\v{s}ev polynomials will suffice.$^{[8,9]}$ In this paper, we will only consider the equation $(24)$.\par When a satellite is exposed to solar radiation, it experiences a force that arises from the absorption or reflection of photons. The resultant acceleration depends on the satellite's mass and surface area. The magnitude of solar radiation pressure is influenced by the solar flux $\phi=\Delta E/A\Delta t$, where  $\Delta E$ is the amount of energy passing through a surface area $A$ during a time interval $\Delta t$. A photon with energy $E_\nu$ carries an impulse $p=E_\nu/c$, where $c$ denotes the velocity of light. Consequently, the overall momentum of an absorbing object, which is exposed to sunlight, alters by\[\Delta p=\frac{\Delta E}{c}=\frac{\phi}{c}A\Delta t\tag{25}\] within the time period $\Delta t$. This suggests that the satellite encounters a force \[F=\frac{\Delta p}{\Delta t}=\frac{\phi}{c}A\tag{26}\] that is directly proportional to the area of cross section $A$. In other words, the satellite is subject to a pressure $P=\phi/c$. In the vicinity of the Earth, at a distance of approximately one astronomical unit (AU) from the Sun, the solar flux $\phi$ measures $1367\,Wm^{-2}$. Assuming the satellite absorbs all incident photons and is perpendicular to the incoming radiation, this leads to a solar radiation pressure of $P_\odot=4.56\times 10^{-6}\,Nm^{-2}$. Subsequently, we will consider an arbitrary orientation of the satellite, taking into account the diffuse reflection of light rather than specular reflection. The surface $A$ is defined by a normal vector $\vec n$, which is inclined at an angle $\theta$ to the vector $\vec e_\odot$ pointing towards the Sun. Assuming that the satellite reflects a fraction $\epsilon$ of the total energy $\Delta E$ it receives and absorbs the remainder $(1-\epsilon)\Delta E$, the resulting force is given by$^{[10]}$ \[\vec F=-P_\odot\cos(\theta)A((1-\epsilon)\vec e_\odot+2\epsilon\cos(\theta)\vec n).\tag{27}\] $\epsilon$ is known as the reflectivity of the material used in satellite construction, and typically ranges from $0.2$ to $0.9$. The distance between an Earth-orbiting satellite and the Sun fluctuates between $147\times10^6\,km$ and $152\times10^6 \,km$ over the course of a year. This is a consequence of the eccentricity of Earth's orbit. This leads to an annual variation of approximately $\pm3.3\%$ in solar radiation pressure.$^{[11]}$ Considering this dependence, we ultimately arrive at the ensuing equation for the acceleration of a satellite resulting from solar radiation pressure. \[\frac{d^2\vec r}{dt^2}=-P_\odot\frac{1}{r^2_\odot}\frac{A}{m}\cos(\theta)((1-\epsilon)\vec e_\odot+2\epsilon\cos(\theta)\vec n),\tag{28}\] where, $m$ represents the mass of the satellite. $\cos\theta=\vec n^T\vec e_\odot$ holds true, as both $\vec n$ and $\vec e_\odot$ are unit vectors. For satellites equipped with extensive solar panel arrays, it is sufficient to assume that the outward-pointing normal vector $\vec n$ aligns with the direction of the Sun. In such cases, equation $(28)$ can be simplified to$^{[1]}$ \[\frac{d^2\vec r}{dt^2}=-P_\odot C_R\frac{A}{m}\frac{\vec r_\odot}{r_\odot^3},\tag{29}\] where $C_R$ represents the radiation pressure coefficient $(1+\epsilon)$.\\ Equation $(29)$ has been derived based on the assumption that the satellite is completely illuminated by the Sun. However, the majority of Earth-orbiting satellites encounter partial or full eclipses when traversing the night side of the Earth. To account for this, a shadow function $\nu$ is introduced, which takes on the value $0$ if the satellite is in the region of total eclipse, $1$ if it is sunlit, and satisfies \[\nu=1-\frac{A}{\pi a^2},\quad a=\arcsin\frac{R_\odot}{|\vec r_\odot-\vec r|}\tag{30}\] if the satellite is in the region of partial eclipse. This generalisation leads to equation for the acceleration experienced by the satellite due to solar radiation pressure.$^{[1,3]}$ \[\frac{d^2\vec r}{dt^2}=-\nu P_\odot C_R\frac{A}{m}\frac{\vec r_\odot}{r_\odot^3}.\tag{31}\] \par Atmospheric forces form the largest component of non-gravitational perturbative forces acting on low-Earth orbiting satellites. The primary atmospheric force, known as drag, acts in the opposite direction of the satellite's velocity relative to the atmospheric flux, thus causing deceleration. Negligible contributions to the atmospheric forces include the lift force and binormal forces, which can be safely disregarded in the majority of cases. Let us examine a small mass element $\Delta m$ within the atmospheric column which intersects the satellite's area of cross section $A$ during a specific time interval $\Delta t$. We have $\Delta m=\rho Av_r\Delta t$. Then, the impulse $\Delta p$ exerted on the satellite is given by $\rho Av_r^2\Delta t$. The acceleration of the satellite caused by atmospheric drag can therefore be formulated as$^{[1,12]}$ \[\frac{d^2\vec r}{dt^2}=-\frac{1}{2}C_D\frac{A}{m}\rho v_r^2\frac{\vec v_r}{v_r}.\tag{32}\] $C_D$ represents the interaction of the atmosphere with the satellite's surface, and is a dimensionless quantity. Typical values for $C_D$ range from $2.0$ to $2.3$ for non-spherical convex shaped spacecraft. $C_D$ is commonly estimated as a free parameter in implementations of orbit determination models. The relative velocity vector $\vec v_r$ can be determined using the expression $\vec v_r=\vec v-\vec w_\oplus\times\vec r$, where $\vec v$ is the satellite's velocity vector, $\vec r$ the position vector, and $\vec w_\oplus$ is the Earth's angular velocity vector, of magnitude $0.7292\times10^{-4}\,rad\,s^{-1}$$^{[11]}$.\par The discussion on atmospheric drag thus far revolves around the assumption that the altitude of the satellite is low. The density of the upper atmosphere depends on various parameters in an intricate manner. Atmospheric density models for the upper atmosphere can be classified into static and dynamic models, the former relying mainly on the altitude of the satellite and the latter considering other factors such as solar flux, Earth's magnetic field, time of day, Earth's magnetic field, and geocentric latitude and longitude of the spacecraft. The Harris-Priester$^{[13]}$ model is a popular static model, and variations of the Jacchia model$^{[14,15]}$ are widely used dynamic models.\\\\
\normalsize\centerline{\textbf{\S3. HIGH PRECISION MODELLING}}
\small\par The accelerations discussed thus far are generally adequate for a broad range of applications. However, certain missions with stringent accuracy demands must consider additional perturbations, including Earth radiation pressure, tidal forces that alter Earth's gravity field, and deviations from the Newtonian equations of motion due to general relativity.\par Apart from the direct solar radiation pressure, the radiation emitted by the Earth exerts a force on the satellite. In contrast to solar radiation pressure, Earth radiation pressure is typically divided into shortwave optical radiation and longwave infrared radiation. In both cases, the acceleration of the satellite slightly decreases as the altitude increases. The solar radiation pressure that reflects off the surface of the central body, which in our case is the Earth, is known as albedo. The magnitude of albedo-induced acceleration for satellites in low-Earth orbits ranges from $10\%$ to $35\%$ of the acceleration resulting from solar radiation pressure. The optical albedo radiation arises from the reflection and scattering of the solar radiation incident on the Earth's surface. This reflected radiation is quantified by the albedo factor $a$, which represents the ratio of shortwave radiation reflected from the Earth's surface back into space to the incoming shortwave solar radiation. The average global albedo value is $a\approx0.34$, and the corresponding radiation from Earth's surface elements amounts to approximately $459\,W\,m^{-2}$.$^{[16]}$ The spectral distribution of the optical albedo radiation closely resembles that of direct solar radiation. It is emitted solely from the illuminated side of the Earth and can exhibit substantial variation due to diverse surface characteristics and cloud cover.\par The other significant form of radiation, infrared radiation, is a nearly isotropic re-emission of solar radiation absorbed by the Earth and its atmosphere. The average emissivity of the Earth surface elements is approximately $\epsilon=0.68$. However, the influence of this emissivity is diminished by a factor of $4$, from $\pi R_\oplus^2/4\pi R_\oplus^2$, the ratio of the total radiating cross section of the Earth to the total irradiated Earth surface. As a result, the effective radiation from Earth surface elements caused by infrared emissions is approximately $230\,Wm^{-2}$.$^{[16,17]}$\par The albedo-induced acceleration of a satellite is aggregated from $N$ individual terms, which arise from the different Earth area elements $dA_j$, and is given by$^{[16]}$ \[\frac{d^2\vec r}{dt^2}=\sum\limits_{j=1}^N C_R\left(\nu_ja_j\cos\theta_j^E+\frac{1}{4}\epsilon_j\right)P_\odot\frac{A}{m}\cos\theta_j^S\frac{dA_j}{\pi r_j^2}\vec e_j.\tag{33}\] In the above equation, $\nu_j$ represents the Earth element shadow functions as described in Equation $(30)$, while $\theta_j^E$ and $\theta_j^S$ denote the angles between the normals of the Earth or satellite surface and the incident radiation, respectively. The direction from the Earth surface element to the satellite, with a distance of $r_j$, is indicated by the unit vector $\vec e_j$. A second-degree zonal spherical harmonic model can be utilised to represent the albedo and the emissivity. Usually, approximately $20$ such Earth surface elements are taken into account.\par The gravitational influence of the Sun and the Moon not only directly affects the satellite, as discussed in section $(2)$, but also exerts forces on the Earth itself. These forces result in time-varying deformations of the Earth, known as solid Earth tides. In addition to solid Earth tides, the oceans respond differently to the gravitational perturbations from the Moon and the Sun, giving rise to ocean tides. Consequently, the Earth's gravitational field exhibits small periodic variations. This in turn affects the motion of Earth orbiting satellites.\par The gravitational field generated by a third body with a mass of $M$, in a co-rotating frame, leads to a potential $U$ at a point $P$ on the Earth's surface. This is expressed as follows.$^{[17,18]}$ \[U=\frac{GM}{|\vec s-\vec R|}+\frac{1}{2}n^2d^2,\tag{34}\] where $\vec R$ and $\vec s$ represent the geocentric coordinate vectors of the point $P$ and the body generating the tides, respectively. Additionally, the average motion of the body around an axis passing through the center of mass of the system is denoted by $n$, and $d$ represents the distance between point $P$ and this axis. For both the Sun and the Moon, $s$ is significantly larger than $R$. Hence, the denominator in $(34)$ can be expanded as follows. \[\frac{1}{|\vec s-\vec R|}\approx\frac{1}{s}\left(1+\frac{R}{s}\cos\gamma-\frac{1}{2}\frac{R^2}{s^2}+\frac{3}{2}\frac{R^2}{s^2}\cos^2\gamma\right),\tag{35}\] where $\gamma$ represents the angle between $\vec s$ and $\vec R$. The distance $d$ can further be expressed as \[d^2=\left(\frac{M_s}{M+M_\oplus}\right)^2+R^2\cos^2\phi-2\frac{M_s}{M+M_\oplus}R\cos\gamma,\tag{36}\] $\phi$ being the geocentric latitude. With the above relations and $n^2s^3= G(M+M_\oplus)$, the geopotential may be rewritten, to convey the contribution of the each individual terms, as$^{[17]}$ \[U=\frac{GM}{s}\left(1+\frac{1}{2}\frac{M}{M+M_\oplus}\right)+\frac{GMR^2}{2s^3}(3\cos^2\gamma-1)+\frac{n^2R^2}{2}\cos^2\phi.\tag{37}\] The first term in $(37)$ remains constant. The rotational potential around an axis passing through the Earth's center and perpendicular to the orbital plane is depicted by the third term. It introduces a minor, permanent bulge around the Earth's equator, resembling that formed by the Earth's rotation, although significantly smaller in magnitude, since $n^2<<\omega_\oplus^2$.\par The tidal potential $U_2$ is the second term in $(37)$. It is a second-order zonal harmonic that deforms the equipotential.$^{[19]}$ Its magnitude is proportional to $GM/s^3$. Therefore, the lunar tides possess about twice the strength of the solar tides. According to the dependence of $U_2$ on $\cos^2\gamma$, the periodicity of the tidal acceleration is predominantly semi-diurnal. The gravitational influence caused by the tides essentially causes the Earth to undergo elastic distortion. This phenomenon can be expressed mathematically as a linear relation between $U_2$ and the resultant altered gravitational potential $U_T$. The Love number $\kappa\approx0.3$ represents the ratio between these two potentials. If the Earth were completely rigid, the Love number would be negligible. Since the tidal potential follows a second-order harmonic pattern, the altered gravitational potential diminishes with $1/r^3$, and can ultimately be represented as$^{[17,18]}$ \[U_T=\frac{1}{2}\kappa\frac{GMR_\oplus^5}{s^3r^3}(3\cos^2\gamma-1).\tag{38}\]\par The perturbations in satellite orbits caused by lunar and solar solid Earth tides are determined by expanding the gravity potential induced by tides using spherical harmonics, similar to the method employed for modeling the static gravity field of the Earth. For practical purposes, the time-varying adjustments to the unnormalized geopotential coefficients can be calculated according to$^{[18]}$ \[\begin{bmatrix}\Delta C_{nm}\\\Delta S_{nm}\end{bmatrix}=4k_n\left(\frac{GM}{GM_\oplus}\right)\left(\frac{R_\oplus}{s}\right)^{n+1}\sqrt{\frac{(n+2)(n-m)!^3}{(n+m)!^3}}\]\[\cdot P_{nm}(\sin\phi)\begin{bmatrix}\cos(m\lambda)\\\sin(m\lambda)\end{bmatrix}\tag{39}\] for the Sun and Moon, where the Love numbers of degree $n$ are represented by $k_n$. The Earth fixed latitude and longitude of the perturbing body are denoted here by $\phi$ and $\lambda$.\par The impact of ocean tides is relatively small when compared to solid Earth tides, typically by approximately one order of magnitude. The effects of ocean tides can be explained by means of an ocean tide potential that is expanded using spherical harmonics and subsequently converted into geopotential coefficients that vary with time.$^{[19]}$ \[\begin{bmatrix}\Delta C_{nm}\\\Delta S_{nm}\end{bmatrix}=\frac{4\pi GR_\oplus^2\rho_w(1+k_n')}{GM_\oplus(2n+1)}\qquad\qquad\qquad\qquad\qquad\qquad\qquad\qquad\]\[\cdot\begin{bmatrix}\sum\limits_{s(n,m)}(C_{snm}^++C_{snm}^-)\cos\theta_s+(S_{snm}^++S_{snm}^-)\sin\theta_s\\\sum\limits_{s(n,m)}(S_{snm}^++S_{snm}^-)\cos\theta_s-(C_{snm}^++C_{snm}^-)\sin\theta_s\end{bmatrix}.\tag{40}\] where the ocean tide coefficients for the tide constituent $s$ are denoted as $C_{snm}^\pm$ and $S_{snm}^\pm$ in meters. The density of ocean water is represented by $\rho_w$, and the $k_n'$ refer to the load deformations. Additionally, $\theta_s$ corresponds to the weighted sum of the six fundamental arguments associated with the orbits of the Sun and Moon.$^{[11]}$\par To provide a comprehensive analysis of the satellite's motion, it is necessary to incorporate the principles of the general theory of relativity. The special theory of relativity assumes a flat four-dimensional space-time, whereas modelling its curvature in the proximity of the Earth requires a different approach. In this context, we utilize the standard coordinates $x^\mu=(ct,x^1,x^2,x^3)$ and examine the invariant element between two events \[ds^2=-c^2d\tau^2=g_{\mu\nu}dx^\mu dx^\nu.\tag{41}\] We can describe the post-Newtonian space-time by expanding these terms to obtain$^{[20]}$ \[ds^2=-\left(1-\frac{2U}{c^2}+\frac{2U^2}{c^4}\right)(dx^0)^2\qquad\qquad\qquad\qquad\]\[-4\frac{V_i}{c^3}dx^0dx^i+\left(1+\frac{2U}{c^2}\delta_{ij}dx^idx^j\right).\tag{42}\]Note that Einstein's summation convention has been applied here, and the indices $\mu,\nu\in\{0,1,2,3\}$, whereas the indices $i,j$ come from the set $\{1,2,3\}$. The times $\tau,t$ represent the time measured by atomic clocks that are moving together with the satellite and positioned at the geocenter, respectively. The gravito-electric effects, described by the first term in equation $(42)$, arise from the curvature of space-time caused by the mass of the Earth, which corresponds to approximately $GM_\oplus/c^2R_\oplus\approx7\times10^{-10}$ at the surface of the Earth. The gravito-magnetic effects, described by the third term in $(42)$, originate from the dragging of space-time caused by the rotation of the Earth, and have a magnitude of approximately $GL_\oplus/c^3R_\oplus^2\approx4\times10^{-16}$.\par The geodesic equation, which governs the motion of a satellite based on the general theory of relativity, is given by \[\frac{d^2x^\mu}{d\tau^2}+\Gamma^\mu_{\nu\sigma}\frac{dx^\nu}{d\tau}\frac{dx^\sigma}{d\tau}=0.\tag{43}\] The Christoffel symbols $\Gamma^\mu_{\nu\sigma}$, which describe fictitious forces arising from a non-inertial frame of reference are obtained through the equation \[\Gamma^\mu_{\nu\sigma}=\frac{1}{2}g^{\alpha\mu}\left(\frac{dg_{\alpha\nu}}{dx^\sigma}+\frac{dg_{\alpha\sigma}}{dx^\nu}-\frac{dg_{\nu\sigma}}{dx^\alpha}\right).\tag{44}\] The geodesic equation can be expanded in relativistic terms $U/c^2$ and $V/c^3$, according to the given metric in the proximity of the Earth. This procedure yields the Newtonian equation of motion with additional correction terms. In the present work, we associate the coordinate time $t$ with the terrestrial time and neglect the contribution of the gravito-magnetic effects to arrive at the post-Newtonian correction to the acceleration$^{[21]}$ \[\frac{d^2\vec r}{dt^2}=-\frac{GM_\oplus}{r^2}\left(\left(4\frac{GM_\oplus}{c^2r}-\frac{v^2}{c^2}\right)\vec e_r+4\frac{v^2}{c^2}(\vec e_r\cdot\vec e_v)e_v\right).\tag{45}\] In the above equation, $\vec e_r$ and $\vec e_v$ represent the unit position and velocity vectors. The relation $GM_\oplus/r=v^2$ holds for circular orbits, where the velocity is orthogonal to the radius vector. Therefore, the relativistic correction to the acceleration \[\frac{d^2\vec r}{dt^2}=-\frac{GM_\oplus}{r^2}\vec e_r\left(3\frac{v^2}{c^2}\right)\tag{46}\] is determined by multiplying the post-Newtonian acceleration with a term $3v^2/c^2\approx3\times10^{-10}$ for a common satellite. As a general guideline, the magnitude of general relativistic effects can be estimated using the Schwarzschild radius of the Earth, which is $2GM_\oplus/c^2\approx1\,cm$. For satellite applications that aim for this level of precision, it is crucial to meticulously account for the effects of general relativity.\\\\
\normalsize\centerline{\textbf{\S4. THE NUMERICAL SOLUTION}}
\small\par The precise level of accuracy demanded by satellite orbit calculations can only be attained by using numerical techniques to solve the mathematical model. In this paper, we use Fehlberg's formulation of the Runge-Kutta class of embedded numerical integrators with adaptive stepsize control. In particular, the RKF-$4(5)$, RKF-$7(8)$, and the RKF-$8(9)$ methods are employed. We will now give an overview of the fixed step Runge-Kutta methods and then proceed to describe Fehlberg's formulation of embedded Runge-Kutta methods.\par We start with the ordinary differential equation $\dot{\vec y}=f(t,\vec y)$, where $\vec y,\dot{\vec y}, f\,\in\,\mathbb{R}^n$. This form can always be obtained from a second order ordinary differential equation $\ddot{\vec x}=a(t,\vec x,\dot{\vec x})$ by combining the position vector $\vec x$ and the velocity vector $\dot{\vec x}$ into a state vector $\vec y=[\vec x\quad\dot{\vec x}]^T$, which satisfies $\dot{\vec y}=f(t,\vec y)$. We start with the initial state vector $\vec y_0=\vec y(t_0)$ and then calculate $\vec y(t_0+h)\approx y_0+h\varphi$, where $\varphi$ is the increment function. The advantage of this method is that it avoids calculating all the derivatives in the Taylor series expansion for $\vec y(t_0+h)$. In a general $n$ stage Runge-Kutta formula, which will have the same accuracy as the approximation by an $n^{th}$ order Taylor polynomial, $n$ function evaluations$^{[22]}$ \[\vec k_1=f(t_0+c_1h+\vec y_0)\]\[\vec k_i=f\left(t_0+c_ih,\vec y_0+h\sum\limits_{j=1}^{i-1}a_{ij}\vec k_j\right),\quad i=2,\cdots,n\tag{47}\] are employed to construct the increment function $\vec\varphi=b_1\vec k_1+b_2\vec k_2+\cdots+b_n\vec k_n$. We then have the approximation \[\vec\eta(t_0+h)=\vec y_0+h\vec\varphi\tag{48}\] to the exact solution $\vec y$. Each Runge-Kutta method is completely specified by its coefficients, typically organized in a Butcher tableau. These coefficients are selected to maximize the order $p$ of the local truncation error, ensuring its accuracy to the highest possible degree.\par An adaptive time stepping formulation may be deduced by means of an embedded pair of Runge-Kutta methods, which yield the two independent estimates$^{[22]}$ \[\hat{\vec\eta}(t_0+h)=\vec y_0+h\sum\limits_{i=1}^n\hat b_i\vec k_i,\quad\vec\eta(t_0+h)=\vec y_0+h\sum\limits_{i=1}^nb_i\vec k_i\tag{49}\] of orders $(p+1)$ and $p$ respectively. Their local truncation errors are bounded by $\hat\epsilon\leq ch^{p+2}$ and $\epsilon\leq ch^{p+1}$ respectively. We can estimate the error $\epsilon\approx|\vec y-\vec\eta|=|\hat{\vec\eta}-\vec\eta|$, because $\hat\epsilon$ is smaller than $\epsilon$ by a order of $h$. This enables us to acquire an approximation of the local truncation error for the formula of order $p$ from the difference $|\hat{\vec\eta}-\vec\eta|$. This is a crucial estimate for adaptive stepsize control.\par When numerically integrating a differential equation, it is essential to select the stepsize in a manner that ensures each step makes a uniform contribution to the overall integration error. The stepsize should obviously not be too large, but it should not be too small either, as this considerably increases round-off errors and computational effort. Assume that a single integration step is performed with a predetermined stepsize $h$, resulting in an estimate of the local truncation error $\epsilon=|\hat{\vec\eta}-\vec\eta|$. We select an error tolerance $\xi$, and if $\epsilon$ exceeds $\xi$, we iterate the process with a reduced stepsize $h^*$. Considering that $\epsilon(h)$ exhibits a directly proportionality to $h^{p+1}$ for a method of order $p$, the local truncation error can be expressed as \[\epsilon(h^*)=\epsilon(h)\left(\frac{h^*}{h}\right)^{p+1}\approx|\hat{\vec\eta}-\vec\eta|\left(\frac{h^*}{h}\right)^{p+1}\tag{50}\] for the new stepsize. By imposing the condition that the local truncation error should be less than $\xi$ and solving for $h^*$, we obtain the largest permissible stepsize for the subsequent iteration$^{[23]}$ \[h^*=\sqrt[p+1]{\frac{\xi}{\epsilon(h)}}=\sqrt[p+1]{\frac{\xi}{|\hat{\vec\eta}-\vec\eta|}}\cdot h.\tag{51}\] In practice approximately $90$ of this maximum allowable stepsize is often employed for safety considerations, to prevent another unsuccessful step. If the current step is successful, we continue with the stepsize $h^*$ for the subsequent iteration. To maintain stability and prevent rapid oscillations in the stepsize, the value of $h$ should not be altered by more than a factor of 2 between consecutive steps.\par Although this kind of stepsize control effectively adjusts the stepsize according to the characteristics of the differential equation, it does not eliminate the requirement that the user has to provide an initial estimate for the starting stepsize. Test calculations and experience are valuable in determining an appropriate initial stepsize for specific problems. For example, when integrating a satellite orbit, one approach could be to begin with a stepsize equal to $1/100^{th}$ of the orbital period, integrating over multiple orbits and observing the calculated step sizes, a suitable initial stepsize can be determined and used for similar calculations. We have provided the Butcher tableaus that have been used in the implementation of the RKF-$4(5)$ and RKF-$7(8)$ algorithms at the end of this section.$^{[23, 24]}$\par In the previous discussion regarding the numerical solution of differential equations of motion, we did not consider the need to obtain the solution at predetermined output points. This does not pose a significant problem as long as the disparity between two consecutive points where the solution is needed is significantly greater than the step size suggested by the adaptive time stepping method. However, in cases where the solution is required at intermediate points, we employ interpolation techniques. In this paper, we have used either Lagrange interpolation or cubic spline interpolation, based on accuracy requirements. We now give an overview of these techniques.\par Given $y=f(x)$ which takes $(n+1)$ values, namely $y_0, y_1,\cdots, y_n$ corresponding to $x_0,x_1,\cdots, x_n$, we wish to construct an interpolating polynomial that approximates the given polynomial $f(x)$, the explicit definition for which is unknown. We have the polynomial approximation$^{[25]}$
$$y=f(x)\approx a_0(x-x_1)(x-x_2)\cdots(x-x_n)$$ $$+a_1(x-x_0)(x-x_2)\cdots(x-x_n)+a_2(x-x_0)(x-x_1)$$
$$\cdots(x-x_n)+\cdots+a_n(x-x_0)(x-x_1)\cdots(x-x_{n-1}),$$
$$\text{where\quad}a_0=\frac{y_0}{(x_0-x_1)(x_0-x_2)\cdots(x_0-x_n)},$$ $$a_1=\frac{y_1}{(x_1-x_0)(x_1-x_2)\cdots(x_1-x_n)},$$
\[\cdots a_k=\frac{y_1}{(x_k-x_0)(x_k-x_1)\cdots(x_1-x_n)}.\tag{52}\]
 We can now consider the Lagrange basis functions $L_i(x)$ to be the following. $$L_0(x)=\frac{(x-x_1)(x-x_2)\cdots(x-x_n)}{(x_0-x_1)(x_0-x_2)\cdots(x_0-x_n)},$$ $$L_1(x)=\frac{(x-x_0)(x-x_2)\cdots(x-x_n)}{(x_1-x_0)(x_1-x_2)\cdots(x_1-x_n)},\cdots$$
\[\cdots,L_n(x)=\frac{(x-x_0)(x-x_2)\cdots(x-x_n)}{(x_k-x_0)(x_k-x_2)\cdots(x_k-x_n)}.\tag{53}\] These are polynomials of degree $(n-1)$, and form the set of basis functions for the Lagrange interpolating polynomial$^{[25]}$
\[p_n(x)=\sum\limits_{i=1}^ny_i\prod\limits_{j=1,\,j\neq i}^n\frac{x-x_j}{x_i-x_j}=\sum\limits_{i=1}^nf(x_i)\prod\limits_{j=1,\,j\neq i}^n\frac{x-x_j}{x_i-x_j}.\tag{54}\]
It now remains to be seen how close the approximation given by this interpolating polynomial is. If a degree $n$ interpolating polynomial is defined over a distinct set of grid points $\{x_0,x_1,..x_n\}$ in the interval $[a,b]$ and $f \in C[a,b]$, then the error in the polynomial is given by the expression
\[E_n=f(x)-p(x)\Rightarrow E_n=\frac{\left((x-x_0)(x-x_1)\cdots(x-x_n)\right)f^{n+1}\xi}{(n+1)!}.\tag{55}\]\par The approximation given by Lagrange interpolation is accurate as long as the degree of the polynomial remains small. But as the degree increases, the number of necessary multiplications increases, making the computation difficult. Moreover, oscillations within the intervals causes the Lagrange interpolating polynomial to be unstable to perturbations. To work around this inconvenience, we sometimes prefer piece-wise interpolation, called spline interpolation. In this paper, we consider cubic spline interpolation.\par Given $n$ data points, we now present the system of equations which has to be solved to obtain the values of $S_i$, which are used to construct the approximating cubic spline in the interval $(x_i,\,x_{i+1})$.$^{[25]}$ \[\begin{bmatrix} h_0&2(h_0+h_1)&h_1&&\\&h_1&2(h_1+h_2)&h_2&&\\&\,&\ddots&\ddots&&\\&\,&h_{n-2}&2(h_{n-2}+h_{n-1})&h_{n-1}\end{bmatrix}\begin{bmatrix}S_0\\S_1\\\vdots\\S_n\end{bmatrix}\]\[=6\begin{bmatrix}f[x_1,x_2]-f[x_0,x_1]\\f[x_2,x_3]-f[x_1,x_2]\\\vdots\\f[x_{n-1},x_{n}]-f[x_{n-2},x_{n-1}]\end{bmatrix}.\tag{56}\] In this paper, we will only consider natural cubic splines, for which $S_0=S_n=0$. Solving $(54)$ yields the approximating cubic polynomial $g_i(x)$ in the interval $(x_i,\,x_{i+1})$. \[g_i(x)=\left(\frac{S_{i+1}-S_i}{6h_i}\right)(x-x_i)^3+\frac{S_i}{2}(x-x_i)^2\qquad\qquad\qquad\quad\]\[+\left(\frac{y_{i+1}-y_i}{h_i}-\frac{2h_iS_i+h_iS_{i+1}}{6}\right)(x-x_i)+y_i.\tag{57}\]
\begin{center}\small
  \includegraphics[width=\linewidth]{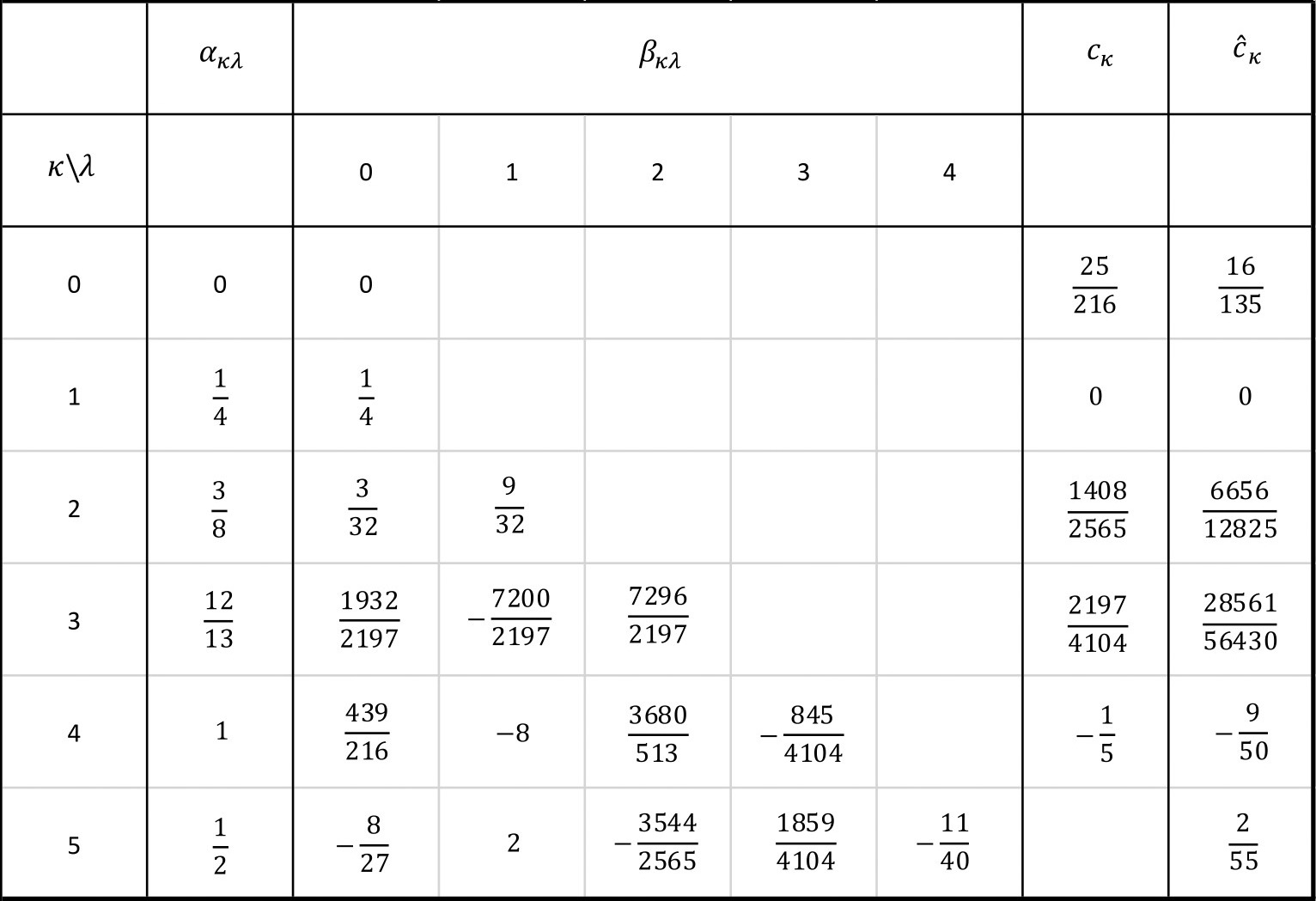}
  \scriptsize\textbf{Table 1.} Butcher Tableau for the RKF-$4(5)$ algorithm
\end{center}
\begin{center}\small
  \includegraphics[width=\linewidth]{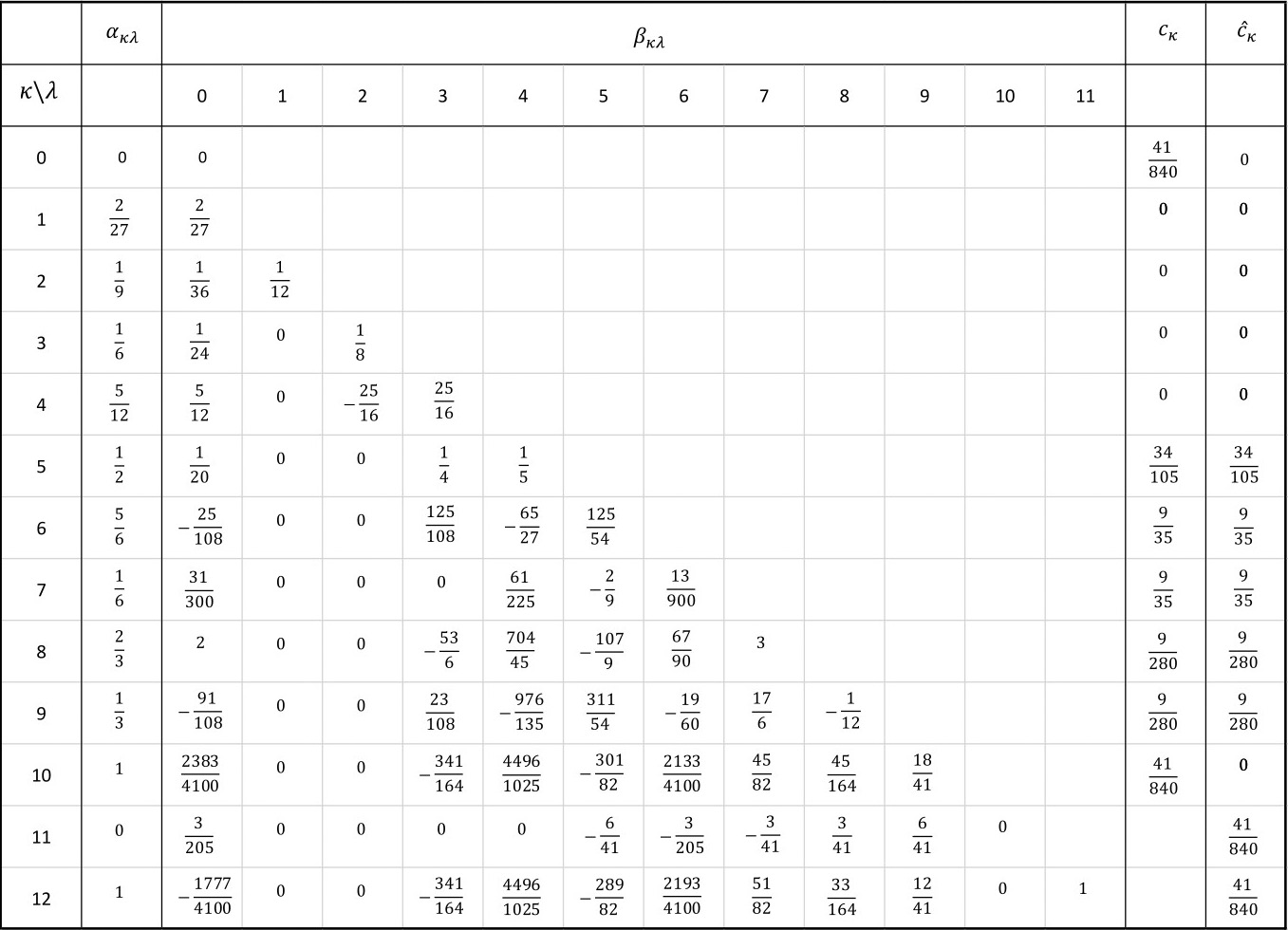}
  \scriptsize\textbf{Table 2.} Butcher Tableau for the RKF-$7(8)$ algorithm\\$\,$
\end{center}
\normalsize\centerline{\textbf{\S5. THE COMPUTATION METHODOLOGY}}
\small\par We have chosen C as the programming language to implement the above mathematical model and the numerical solution. It has the advantage of executing quickly and being easy to debug. We call the program SED, the Satellite Ephemeris Determiner. SED can predict the satellite ephemeris given the initial state vector of the satellite and the ephemerides of the Sun and the Moon in the ECI frame of reference. The conversion of the state vector and the input ephemerides to the Earth-Centered Earth-Fixed (ECEF) frame of reference, together with the computation of the orbital elements is performed by SED prior to the calculation of the acceleration of the satellite. The output of the program is also given in the ECI frame.\par We now provide a more comprehensive breakdown of the methodology employed in the implementation. SED takes the initial six-dimensional state vector of the satellite, the ephemerides of the Sun and the Moon, and the surface area of the satellite as the input. The conversion of the state vector to orbital elements is necessary for further computations. The set of six orbital parameters necessary to fully characterize the trajectory of a satellite are the eccentricity of the orbit $(e)$, the length of the semi-major axis $(a)$, the inclination of the orbit $(i)$, the longitude of the ascending node $(\Omega)$, the argument of the periapsis $(\omega)$, and the true anomaly $(\nu)$. The conversion of the state vector to the orbital elements is performed through the following procedure.$^{[3,12]}$
\par The areal velocity vector $\vec h$ is computed from the initial state vector $[x\,\,y\,\,z\,\,\dot x\,\,\dot y\,\,\dot z]$ using the equation \[\vec h=\begin{bmatrix}y\dot z-z\dot y\\z\dot x-x\dot z\\x\dot y-y\dot x\end{bmatrix}.\tag{58}\] The modulus of $\vec h$ is then obtained, using which we denote the normalised vector $\vec W=\vec h/h$. From the representation of $\vec W$ in terms of $i$ and $\Omega$, it follows that$^{[1]}$ \[\begin{bmatrix}W_x\\-W_y\\W_z\end{bmatrix}=\begin{bmatrix}\sin i\,\sin\Omega\\\sin i\cos\Omega\\\cos i\end{bmatrix}.\tag{59}\]
  Hence, we have the following expressions for the inclination of the orbit and the longitude of the ascending node. \[i=\arctan\left(\frac{\sqrt{W_x^2+W_y^2}}{W_z}\right),\quad\Omega=\arctan\left(\frac{W_x}{-W_y}\right).\tag{60}\] The areal velocity can further be used to deduce the semi-latusrectum $p=h^2/GM_\oplus$. Next, using the vis-viva equation, we obtain the length of the semi-major axis$^{[26]}$ \[a=\left(\frac{2}{r}-\frac{v^2}{GM_\oplus}\right)^{-1},\tag{61}\] and consequently, the mean motion of the satellite \[n=\sqrt{\frac{GM_\oplus}{a^3}}.\tag{62}\] In the case of elliptical orbits, the value of $a$ will always be positive. The eccentricity $e$ can then be determined from the following equation. \[e=\sqrt{1-\frac{p}{a}}.\tag{63}\] We now solve for the eccentric anomaly $E$ using the equation \[E=\arctan\left(\frac{\vec r\cdot\dot{\vec r}/(a^2n)}{1-r/a}\right),\tag{64}\] whence the true anomaly is given by$^{[1,3]}$ \[\nu=\arctan\left(\frac{\sqrt{1-e^2}\sin E}{\cos E-e}\right).\tag{65}\] To calculate the argument of the periapsis $\omega$, we first determine the argument of the latitude $u$ using the equation \[u=\arctan\left(\frac{z}{-xW_y+yW_x}\right).\tag{66}\] We finally compute $\omega$ using\[\omega=u-\nu.\tag{67}\] Using the six orbital elements $e,\,a,\,i,\,\Omega,\,\omega$, and $\nu$, we obtain the geocentric latitude and longitude of the satellite using the equation \[\begin{bmatrix}\sin\phi\\\tan(\lambda-\Omega)\end{bmatrix}=\begin{bmatrix}\sin i\sin(\omega+\nu)\\\cos i\tan(\omega+\nu)\end{bmatrix}.\tag{68}\] Using the geocentric latitude $\lambda$ and longitude $\phi$, we compute the acceleration of the satellite due to Earth's geopotential using $(11)$. In this paper, the geopotential coefficients have been taken from the Joint Gravity Model (JGM-$3$).$^{[6,7]}$ This is taken to be the fundamental acceleration of the satellite obtained with respect to the Newtonian two-body problem. The perturbation accelerations will now be evaluated and added to the acceleration due to the geopotential.\par The largest force acting on the satellite, apart from the central body gravitational force, is third body gravitation. The third bodies of interest here are the Sun and the Moon, whose ephemerides have been taken as inputs to SED. These ephemerides can be obtained with a high degree of accuracy through the use of the JPL SPICE NAIF toolkit.$^{[27]}$ The acceleration of the satellite due to third body gravitational effects is now computed using $(24)$.\par Computing the acceleration of the satellite due to solar radiation pressure (SRP) requires the knowledge of the initial satellite state vector, the geocentric position vector of the Sun, and the area of the satellite exposed to the Sun. For most satellites, we may assume that the outward pointing normal points in the direction of the Sun, so that the surface area of the satellite can be taken to be the area exposed to the Sun. With these input values, the acceleration of the satellite due to SRP can be calculated using $(31)$. The orbital perturbations resulting from shadow transits is accounted for in $(31)$ by means of a shadow function $\nu$. Determining whether the satellite is in the umbra, the penumbra, or the fully sunlit region of its orbit is necessary in the implementation of the shadow function, and is done as follows.$^{[17]}$
\begin{center}\small
  \includegraphics[width=\linewidth]{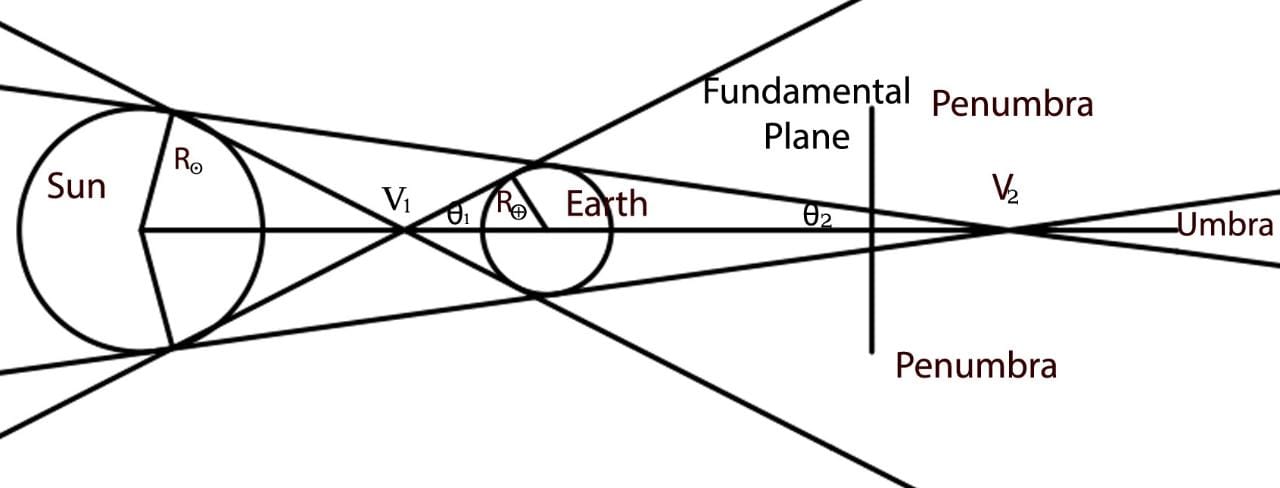}
  \scriptsize\textbf{Figure 1.} Schematic Diagram of the Conical Shadow Model
\end{center}
\par The computation of the extent to which the Sun is obscured by a celestial object such as the Earth is determined by considering the angular separation and diameters of the two bodies. Since the Sun has a relatively small apparent diameter, it is satisfactory to represent the occultation by overlapping circular discs. As we have already stated in $(30)$, we compute the parameter $a$ to calculate the shadow function $\nu$. Similarly, we now define two more parameters. Let$^{[17]}$ \[b=\arcsin\frac{R_\oplus}{r},\quad c=\frac{-{\vec r}^T(\vec r_\odot-\vec r)}{r|\vec r_\odot-\vec r|}.\tag{69}\] If $a+b\leq c$, no occultation takes place, and $\nu=1$. If $c<b-a$, the occultation is total, and $\nu=0$. If $|a-b|<c<a+b$, then the occultation is partial, and we use $(30)$ to compute $\nu$.\par In computing the deceleration of the satellite due to atmospheric drag, we make a distinction between low-Earth orbit satellites (LEO satellites) and satellites orbiting the Earth in the upper atmosphere. For LEO satellites, the state vector of the satellite in the J$2000$ ECI frame of reference is used to compute the atmospheric drag using $(32)$. For satellites orbiting Earth in the upper atmosphere, we use either the static Harris-Priester density model$^{[13]}$ or the dynamic Jacchia density model$^{[15]}$ to obtain the density of the upper atmosphere. The deceleration of the satellite due to atmospheric drag is then calculated with reference to this density. Note that geostationary satellites orbiting at a radius of around $35,786\,km$ from the surface of the Earth experience no atmospheric drag, because the upper atmosphere only extends to about $10,000\,km$ from the surface of the Earth.\par After accounting for these perturbative forces in the differential equation of motion, we check if the user requires a high-precision satellite ephemeris. For most applications, the above computed perturbations suffice in terms of accuracy requirements. If this is the case, we proceed to solve the differential equation of motion we have obtained using a numerical integrator. However, some applications such as satellite geodesy have challenging accuracy requirements, for which we account for three more forces, namely Earth radiation pressure, solid Earth tides and ocean tides, and general relativistic perturbation.\par The perturbation acceleration due to Earth radiation pressure, or the acceleration due to the albedo of the Earth, is summed up from the contributions of different Earth elements. We consider $N=20$ such elements. The acceleration due to each element is computed using the expression within the summation in $(33)$.\par The practical calculation of the perturbative effects brought about by solid Earth tides and ocean tides is a complex undertaking, and requires a prior calculation of the tidal coefficients that represent the gravitational potential variations caused by solid Earth tides and ocean tides. In this work, we have chosen the coefficients according to $(39)$ for lunisolar solid Earth tides, and according to $(40)$ for ocean tides.$^{[18,19]}$ The tidal acceleration experienced by the satellite is now computed by correcting the geopotential in $(15)$ using the modified coefficients.\par The final perturbation that needs to be accounted for in a high-precision model is due to the curvature of space-time around Earth. In general, this need only be accounted for if the application demands a level of accuracy that is of the order of the Schwarzschild radius of the Earth, which is about $1\,cm$. If this degree of accuracy is required, then a relativistic correction to the predicted ephemeris can be made through $(45)$.$^{[20,21]}$
\par Once the total perturbed acceleration has been obtained, SED has to be supplied with the user's selection of the numerical integrator. The choice of numerical integrators provided by SED includes fixed and adaptive time step controlled RKF-$4(5)$, RKF-$7(8)$, and RKF-$8(9)$ methods. Fehlberg's formulation of the Runge-Kutta methods is used for the numerical integration process, as shown in $(49)$ to $(51)$. The Butcher tableau for the chosen integrator is then set up, and SED computes the predicted satellite ephemeris in the J$2000$ ECI frame of reference. If the ephemerides at an intermediate time are required, either Lagrange interpolation or cubic spline interpolation is used to obtain the necessary values.$^{[24,25]}$\\\\
\normalsize\centerline{\textbf{\S6. THE IMPLEMENTATION AND RESULTS}}
\small\par In this section, we present the architecture of SED, and the results it generates. We compare these results with those obtained from the High Precision Orbit Propagator of the Systems Tool Kit (HPOP/STK). Given below is the schematic computational process diagram of SED for the ephemeris determination of Earth orbiting satellites.
\begin{center}
  \includegraphics[width=\linewidth]{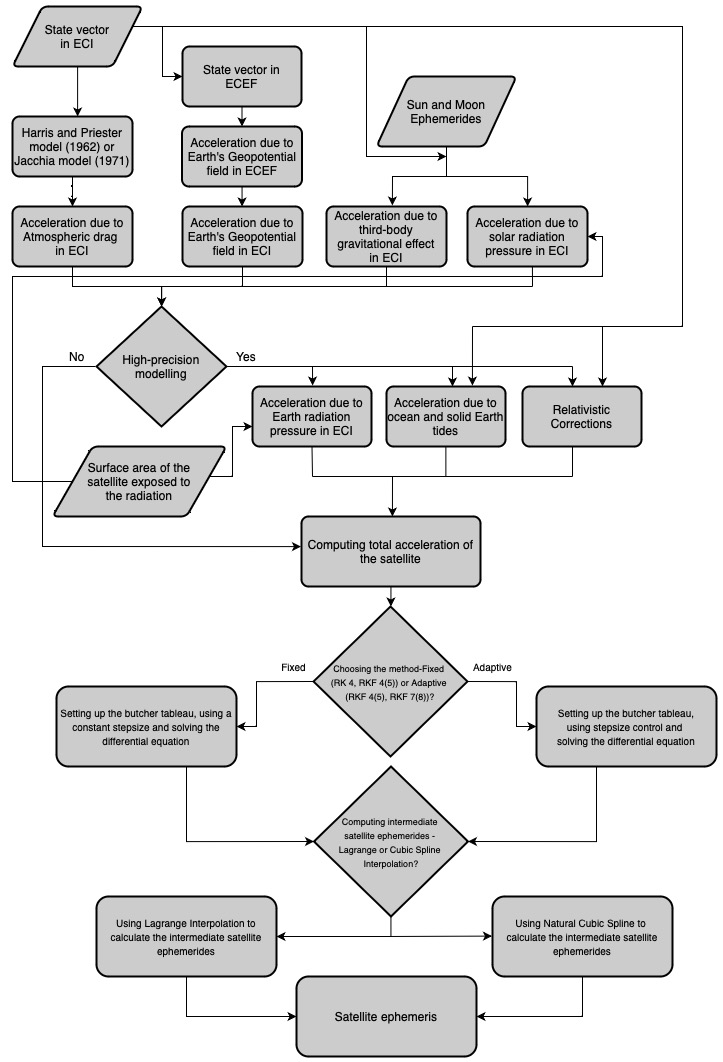}
  \scriptsize\textbf{Figure 2.} Schematic Computational Process Diagram of SED for\\ Ephemeris Determination of Earth Orbiting Satellites
\end{center}
\par We validate the output produced by SED with the results from HPOP/STK. We first ensure that the Runge-Kutta-Fehlberg numerical integrator is functioning accurately by comparing the unperturbed ephemeris predicted by SED for the Indian Remote Sensing (IRS) satellite RESOURCESAT-$2$, which has an orbit altitude of about $850\,km$, with that of HPOP/STK.
\begin{center}
  \includegraphics[width=\linewidth]{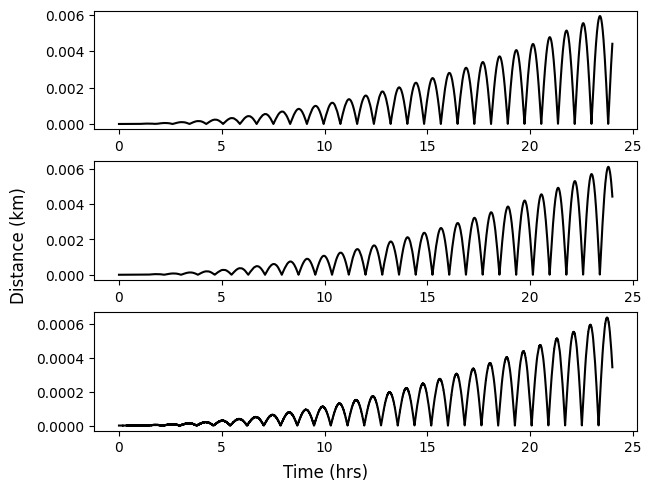}
  \scriptsize\textbf{Figure 3.} SED vs. HPOP/STK error plots for $x, y$, and $z$ coordinate unperturbed position ephemerides respectively, RESOURCESAT-$2$, using RKF-$7(8)$
\end{center}
\begin{center}\small
  \includegraphics[width=\linewidth]{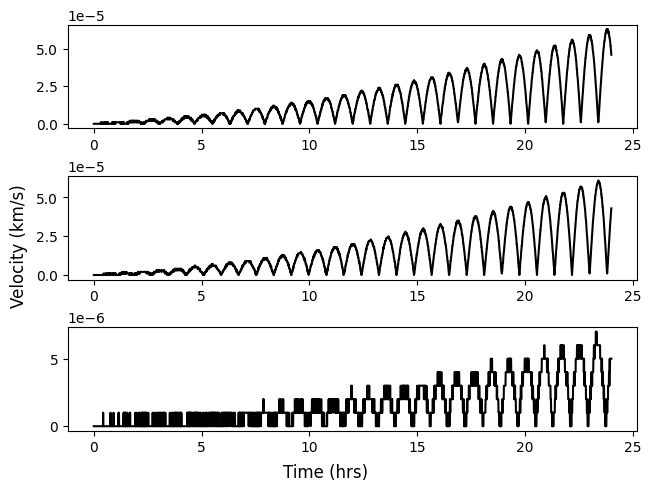}
  \scriptsize\textbf{Figure 4.} SED vs. HPOP/STK error plots for $x, y$, and $z$ coordinate unperturbed velocity ephemerides respectively, RESOURCESAT-$2$, using RKF-$7(8)$
\end{center}
\small The difference between the unperturbed position and velocity ephemerides predicted by SED and HPOP/STK following one day of propagation is summarised in the table presented below.
\begin{center}\small
  \includegraphics[width=\linewidth]{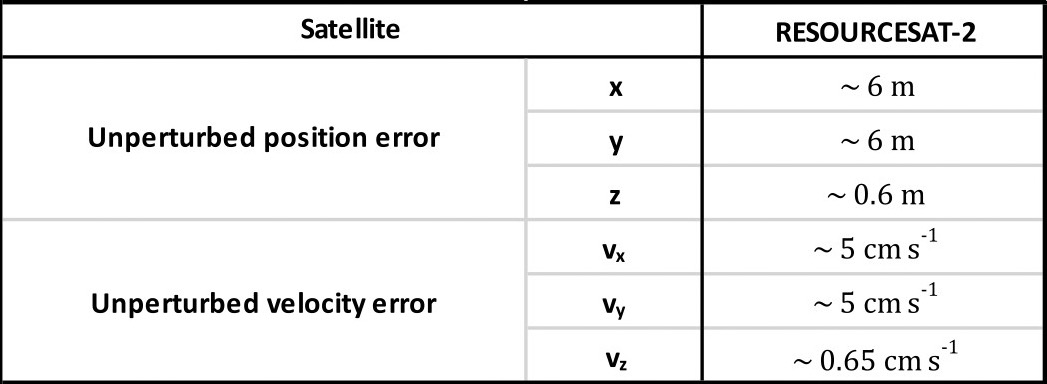}
  \scriptsize\textbf{Table 3.} RESOURCESAT-$2$, Error data using RKF-7(8)
\end{center}
\small The very small amplitude of the error shows us that the numerical integrator is functioning accurately. Given below is the plot illustrating the variation in the stepsize proposed by the RKF-$7(8)$ integrator.
\begin{center}\small
  \includegraphics[width=\linewidth]{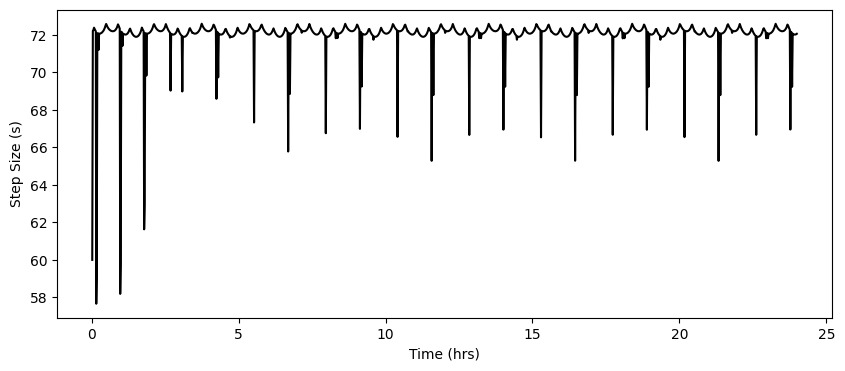}
  \scriptsize\textbf{Figure 5.} Stepsize variation in ephemeris determination, RESOURCESAT-$2$, using RKF-$7(8)$
\end{center}
We can see how the integrator automatically recognises the accumulation of integration errors when the satellite traverses the periapsis of the orbit, and reduces the stepsize. Once the satellite approaches the apoapsis, the integrator once again recognises that the error is small, and increases the step size to maintain computational efficiency.
\par We now compare the perturbed ephemeris generated by SED for two SPACE-X satellites, namely STARLINK-$5466$ and STARLINK-$5458$, which orbit the Earth at an approximate altitude of about $550\,km$, with that of HPOP/STK. Presented below are the plots depicting the difference between SED and HPOP/STK in the predicted perturbed ephemerides.
\begin{center}
  \includegraphics[width=\linewidth]{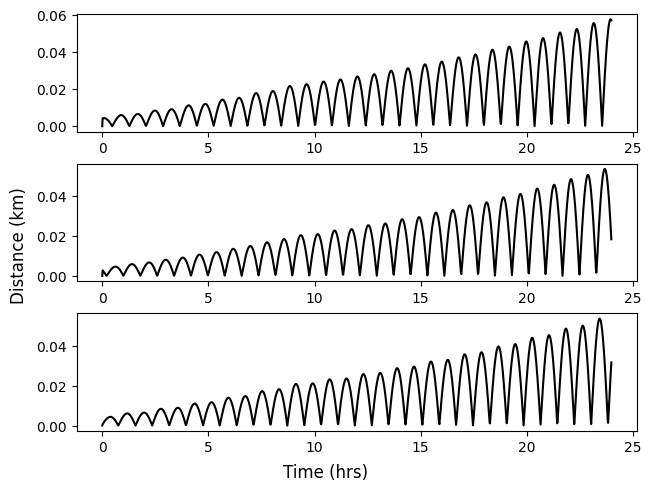}
  \scriptsize\textbf{Figure 6.} SED vs. HPOP/STK error plots for $x, y$, and $z$ coordinate perturbed position ephemerides respectively, STARLINK-$5466$, using RKF-$7(8)$\\
  \includegraphics[width=\linewidth]{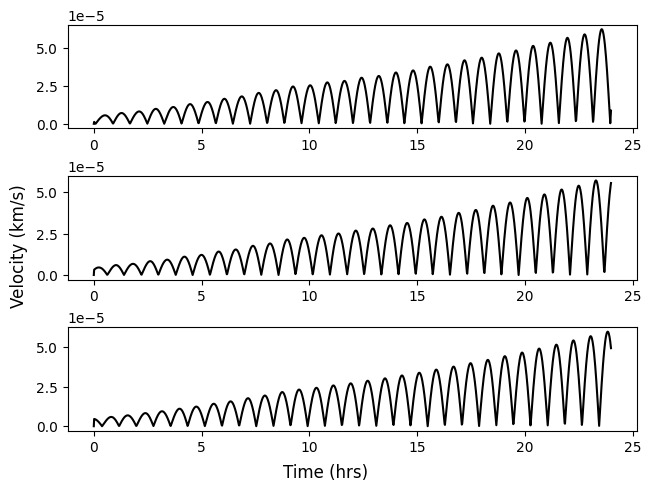}
  \scriptsize\textbf{Figure 7.} SED vs. HPOP/STK error plots for $x, y$, and $z$ coordinate perturbed velocity ephemerides respectively, STARLINK-$5466$, using RKF-$7(8)$
\end{center}
\begin{center}
  \includegraphics[width=\linewidth]{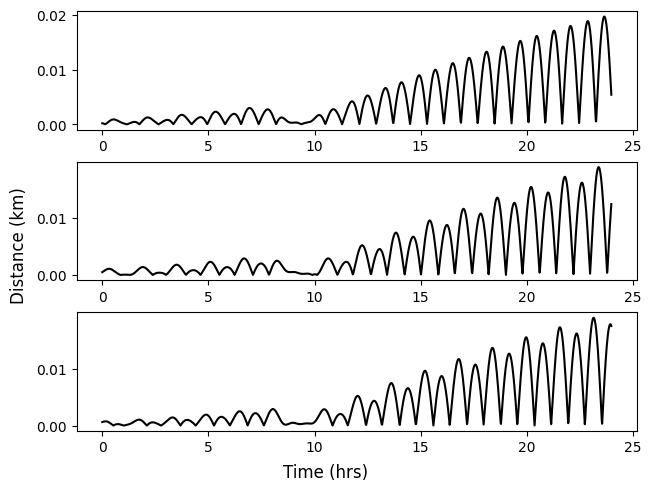}
  \scriptsize\textbf{Figure 8.} SED vs. HPOP/STK error plots for $x, y$, and $z$ coordinate perturbed position ephemerides respectively, STARLINK-$5458$, using RKF-$8(9)$\\
  \includegraphics[width=\linewidth]{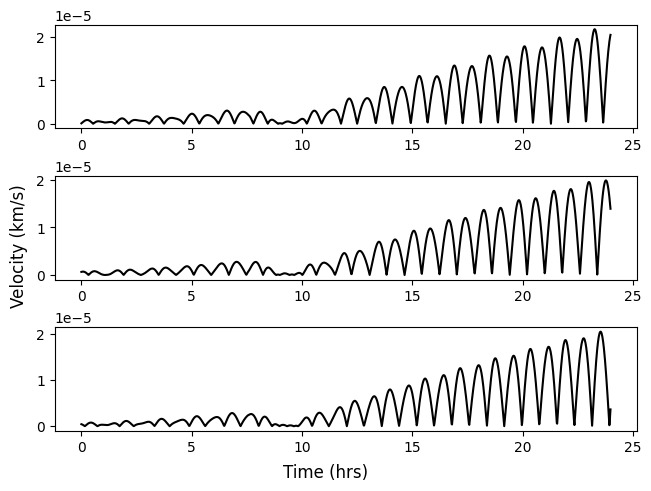}
  \scriptsize\textbf{Figure 9.} SED vs. HPOP/STK error plots for $x, y$, and $z$ coordinate perturbed velocity ephemerides respectively, STARLINK-$5458$, using RKF-$8(9)$
\end{center}
\small Presented below is the table containing information about the different models used in the implementation and the difference in the perturbed position and velocity ephemerides predicted by SED and HPOP/STK following one day of propagation.
\begin{center}
  \includegraphics[width=\linewidth]{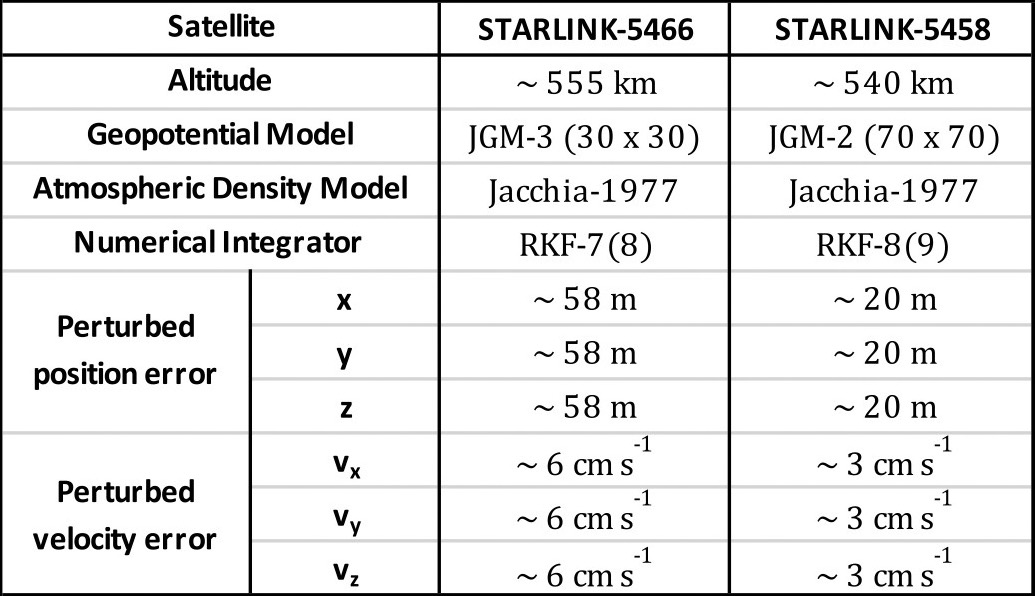}
  \scriptsize\textbf{Table 4.} STARLINK-$5466$ and STARLINK-$5458$, error data using RKF-$7(8)$ and RKF-$8(9)$ respectively
\end{center}
\small Once again, we observe that the perturbed model generates a satisfactory output, exhibiting a minimal margin of error.\\\\
\normalsize\centerline{\textbf{\S7. CONCLUSION}}
\small\par The interest in developing satellite orbit determination models and orbit propagators is primarily being motivated by the need to forecast the future positions of the numerous satellites orbiting the Earth. In this paper, we have formulated a satellite orbit determination model to address this problem. We have formulated the problem in the form of the Newtonian two-body problem, choosing the acceleration due to the geopotential of the Earth to be the fundamental central body acceleration. The model also incorporates the perturbing effect of forces such as third body gravitation, atmospheric drag, and solar radiation pressure. When a high level of precision is required, we have also accounted for additional forces, including solid Earth tides, ocean tides, Earth's albedo, and relativistic perturbations. Explicit formulas have been provided for the calculation of the effect of each force.\par The computation of certain terms in the mathematical expressions describing the perturbation accelerations sometimes poses practical challenges. In such situations, we have provided recurrence relations to aid implementation. Solving the model to obtain the satellite ephemeris requires us to numerically integrate the differential equation of motion, for which we have utilized Fehlberg's formulation of the embedded Runge-Kutta class of integrators with adaptive stepsize control. In particular, the RKF-$4(5)$, RKF-$7(8)$, and RKF-$8(9)$ methods have been used. To obtain the ephemerides at necessary intermediate points, we have used either Lagrange interpolation or cubic spline interpolation.\par We have implemented the mathematical model together with the numerical integrator to obtain a high-fidelity orbit propagator, which we have called the Satellite Ephemeris Determiner (SED). SED employs an intricate computational methodology to maintain a balance between precision and computational efficiency. The set of six orbital elements is computed by SED from the state vector of the satellite at every iteration. We have used the Joint Gravity Model (JGM-$3$) for the central body force model and the Jacchia-1977 model for atmospheric density. When precise ephemerides for the Sun and the Moon are required, we have obtained these using the JPL SPICE NAIF toolkit.\par The current most popular orbit propagator is the High Precision Orbit Propagator of the Systems Tool Kit (HPOP/STK). We have validated the unperturbed and perturbed ephemeris generated by SED with HPOP/STK, and have shown that the magnitude of the difference between the SED ephemerides and HPOP/STK ephemerides is very small. In making this comparison, we have generated the position and velocity ephemerides for three satellites, namely the Indian Remote Sensing (IRS) satellite RESOURCESAT-$2$, and two SPACE-X satellites STARLINK-$5466$ and STARLINK-$5458$.\par SED is currently implemented for the two-body problem, and further work will also enable it to handle the three-body problem. Future research can concentrate on two main areas. The first is to gain a deeper understanding and to accurately model the satellite orbit and its surrounding environment. The second is to devise new numerical techniques that mitigate the accumulation of integration errors. One interesting research direction is the utilization of radial basis function collocation methods to solve the perturbed three-body problem.\\\\
\normalsize\centerline{\textbf{\S8. BIBLIOGRAPHY}}\\\\
\small$[1]\quad$Montenbruck and Gill $(2000)$, \emph{Satellite Orbit Models, Methods, and Applications}, Springer Verlag.\\
$[2]\quad$D.A. Vallado $(2001)$, \emph{Fundamentals of Astrodynamics and Applications}, Springer Verlag, $2$\ts{nd} edition.\\
$[3]\quad$D.A. Vallado $(2013)$, \emph{Fundamentals of Astrodynamics and Applications}, Springer Verlag, $5$\ts{th} edition.\\
$[4]\quad$Herbert Goldstein $(1980)$, \emph{Classical Mechanics}, Addison-Wesley Publishing House, $2$\ts{nd} edition.\\
$[5]\quad$Abramowitz and Stegun $(1970)$, \emph{NIST handbook of Mathematical Functions}, Cambridge University Press.\\
$[6]\quad$Tapley et al $(1996)$, \emph{The Joint Gravity Model JGM-$3$}, American Geophysical Union.\\
$[7]\quad$L.E. Cunningham $(1970)$, \emph{Celestial Mechanics}, Springer Verlag.\\
$[8]\quad$X.X. Newhall $(1993)$, \emph{Ephemerides}, NASA Technical Report.\\
$[9]\quad$P.K. Seidelmann $(1992)$, \emph{Explanatory Supplement to the Astronomical Almanac}, American Astronomical Society.\\
$[10]\quad$Marshall et al $(2007)$, \emph{A Two-Season Impact Study of Satellite Data in the NCEP Global Data Assimilation System}, American Meteorological Society.\\
$[11]\quad$McCarthy $(1996)$, \emph{IERS Conventions}, IERS Technical Note.\\
$[12]\quad$A. de Laco Veris $(2018)$, \emph{Practical Astrodynamics}, Springer Verlag, $1$\ts{st} edition.\\
$[13]\quad$Harris and Priester $(1963)$, \emph{Relation Between Theoretical and Observational Models Of The Upper Atmosphere}, Springer Verlag.\\
$[14]\quad$L.G. Jacchia $(1971)$, \emph{Revised Static Models of the Thermosphere and Exosphere with Empirical Temperature Profiles}, NASA Technical Report.\\
$[15]\quad$L.G. Jacchia $(1977)$, \emph{Thermospheric Temperature, Density, and Composition: New Models}, SAO special report.\\
$[16]\quad$Knocke et al $(1989)$.\\
$[17]\quad$Bertotti and Farinella $(1990)$, \emph{Physics of the Earth and the Solar System}, NASA Technical Report.\\
$[18]\quad$Sanchez $(1994)$, \emph{A Geopotential Model from Satellite Tracking, Altimeter, and Surface Gravity Data: GEM-T$3$}, American Geophysical Union.\\
$[19]\quad$Eanes et al $(1983)$, \emph{Secular Variation of Earth's Gravitational Harmonic J$2$ Coefficient from Lageos and Non-Tidal Acceleration of Earth's Rotation}, Nature Journal.\\
$[20]\quad$Soffel $(1989)$, \emph{Relativity in Astrometry, Celestial Mechanics, and Geodesy}, Springer Verlag.\\
$[21]\quad$Steven Weinberg $(1972)$, \emph{Gravitation and Cosmology: Principles and Application of the General Theory of Relativity}, John Wiley and Sons Inc.\\
$[22]\quad$Stoer and Bulirsch $(1993)$, \emph{Introduction to Numerical Analysis}, Springer Verlag.\\
$[23]\quad$Erwin Fehlberg $(1968)$, \emph{Classical Fifth-, Sixth-, Seventh-, and Eighth-Order Runge-Kutta Formulas with Adaptive Stepsize Control}, NASA Technical Report.\\
$[24]\quad$J.C. Butcher $(1963)$, \emph{Coefficients for the Study of Runge-Kutta Integration Processes}, Australian Mathematical Society.\\
$[25]\quad$Kincaid and Cheney $(2009)$, \emph{Numerical Analysis, Mathematics of Scientific Computing}, American Mathematical Society, $3$\ts{rd} edition.\\
$[26]\quad$V. de Wood $(1878)$, \emph{Force, Momentum, and Vis-Viva}, van Nostrand's Magazine, $18$\ts{th} volume, $109$\ts{th} issue.\\
$[27]\quad$Bachman et al. $(2017)$, \emph{A Look Towards the Future of Handling Space Science Mission Geometry}, NASA Planetary and Space Science Journal.\\
\end{document}